\documentclass[galaxies,conferencereport,accept,moreauthors,pdftex,10pt,a4paper]{mdpi} 
\firstpage{1} 
\articlenumber{33}
\doinum{10.3390/galaxies5030033}
\pubvolume{5}
\pubyear{2017}
\copyrightyear{2017}
\externaleditor{Academic Editors: Duncan A. Forbes and Ericson D. Lopez}
\history{Received: 30 June 2017; Accepted: 28 July 2017; Published: 2 August 2017}

\usepackage{booktabs} 
\usepackage{multirow}
\usepackage{soul} 
\usepackage{microtype}

\usepackage{aas_macros}
\def\eqsim{\mathrel{\raise0.35ex\hbox{$\scriptstyle =$}\kern-0.6em
    \lower0.40ex\hbox{{$\scriptstyle \sim$}}}}
\def\gtrsim{\mathrel{\raise0.35ex\hbox{$\scriptstyle >$}\kern-0.6em
    \lower0.40ex\hbox{{$\scriptstyle \sim$}}}}
\def\lesssim{\mathrel{\raise0.35ex\hbox{$\scriptstyle <$}\kern-0.6em
    \lower0.40ex\hbox{{$\scriptstyle \sim$}}}}
\def\apostle{{\sc apostle}}

\Title{The ``Building Blocks'' of Stellar Halos}

\Author{{Kyle A. Oman}   
$^{1,}$*, Else Starkenburg $^{2}$ and Julio F. Navarro $^{1,\dagger}$}

\AuthorNames{K. A. Oman, E. Starkenburg and J. F. Navarro}

\address{%
$^{1}$ \quad Department of Physics \& Astronomy, University of Victoria, Victoria, BC V8P 5C2, Canada; jfn@uvic.ca\\
$^{2}$ \quad Leibniz Institute for Astrophysics Potsdam (AIP), An der Sternwarte 16, 14482 Potsdam, Germany; estarkenburg@aip.de
}

\corres{Correspondence: koman@uvic.ca; Tel.: +1-250-721-7747}

\firstnote{Senior CIfAR fellow.} 
\abstract{The stellar halos of galaxies encode their accretion histories. In particular, the median metallicity of a halo is determined primarily by the mass of the most massive accreted object. We use hydrodynamical cosmological simulations from the \apostle\ project to study the connection between the stellar mass, the metallicity distribution, and the stellar age distribution of a halo and the identity of its most massive progenitor. We find that the stellar populations in an accreted halo typically resemble the old stellar populations in a present-day dwarf galaxy with a stellar mass $\sim$$0.2$--$0.5$~dex greater than that of the stellar halo. This suggests that had they not been accreted, the primary progenitors of stellar halos would have evolved to resemble typical nearby dwarf irregulars.
}


\begin{document}

%%%%%%%%%%%%%%%%%%%%%%%%%%%%%%%%%%%%%%%%%%
%% Only for the journal Gels: Please place the Experimental Section after the Conclusions

%%%%%%%%%%%%%%%%%%%%%%%%%%%%%%%%%%%%%%%%%%

\section{Introduction}

The accretion of smaller systems is an integral part of galaxy formation. The accretion history of a galaxy is perhaps most clearly encoded in its stellar halo, due to the combination of a relative scarcity of stars formed ``in-situ'', and long dynamical times which allow orbital information to persist. Motivated by the large apparent differences in mass and metallicity between the stellar halos of the Milky Way (MW) and M~31, \cite{2005MNRAS.363L..16R,2006ApJ...646..886F} proposed a picture---supported by cosmological simulation work---in which the metallicity of accreted material reflects the assembly history of the galaxy. The median metallicity of a stellar halo is now thought to be a reflection of the mass of the most massive accreted object \cite{2005ApJ...632..872R,2016ApJ...821....5D,2017arXiv170508442D}, a notion supported by the recent first observation of the stellar halo mass--metallicity relation \cite{2017MNRAS.466.1491H}.

Below, we propose a simple method to explore, within a theoretical framework, the possible link between the accreted halo of a galaxy and present-day dwarf galaxies which may resemble those disrupted to form it.

%%%%%%%%%%%%%%%%%%%%%%%%%%%%%%%%%%%%%%%%%%

\section{Materials and Methods}

We use the \apostle\footnote{A Project Of Simulating The Local Environment}~suite of cosmological hydrodynamical simulations \cite{2016MNRAS.457.1931S,2016MNRAS.457..844F}. These comprise twelve volumes, each containing two halos with masses, separations, kinematics, and local environment consistent with the MW, M~31, and the Local Group of galaxies. Each volume is simulated at mutliple resolution levels, with gas particle masses varying from $\sim$$10^6\,{\rm M}_\odot$ at the lowest (L3) resolution level to $\sim$$10^4\,{\rm M}_\odot$ at the highest (L1) level. In this study, we use the intermediate (L2) level with gas particles of $\sim$$10^5\,{\rm M}_\odot$, dark matter particle mass $\sim$$6\times 10^5\,{\rm M}_\odot$, and $\sim$$300\,{\rm pc}$ force softening. This is the highest resolution at which all twelve volumes have been integrated to the present day. Each volume samples a region extending to radii $\gtrsim$$2\,{\rm Mpc}$ around the barycentre of the two central objects. We assume the WMAP7 cosmological parameters \cite{2011ApJS..192...18K}.

\apostle\ {uses}  
the same hydrodynamics and galaxy formations prescriptions as the {\sc {eagle}}  
project~\cite{2015MNRAS.446..521S,2015MNRAS.450.1937C}---specifically, the model labelled ``Ref'' by \cite{2015MNRAS.446..521S}. The hydrodynamics are solved using the pressure--entropy formulation of smoothed particle hydrodynamics \cite{2013MNRAS.428.2840H}, and the {\sc {anarchy}} collection of numerical methods (for a brief description, see \cite{2015MNRAS.446..521S}) is used. The model includes prescriptions for radiative cooling \cite{2009MNRAS.393...99W}, star formation \cite{2004ApJ...609..667S,2008MNRAS.383.1210S}, stellar and chemical enrichment \cite{2009MNRAS.399..574W}, stellar feedback \cite{2012MNRAS.426..140D}, and cosmic reionization \cite{2001cghr.confE..64H,2009MNRAS.399..574W}. The model is calibrated to reproduce the galaxy mass--size relation and galaxy stellar mass function of $M_\star > 10^8\,{\rm M}_\odot$ objects \cite{2015MNRAS.450.1937C}.

Structures are identified in the simulation output using the friend-of-friends (FoF) \cite{1985ApJ...292..371D} and {\sc {subfind}} \cite{2001MNRAS.328..726S,2009MNRAS.399..497D} algorithms. The former iteratively links particles separated by less than $0.2\times$ the mean inter-particle separation; the latter then identifies self-bound substructures, termed ``subhalos'', separating them along saddle points in the density distribution. The most massive object in each FoF group is labelled the ``central'' object, and other objects in the same group are ``satellites''.

In this study we focus on the stellar halo component of the $24$ roughly MW- and M~31-like galaxies (two per volume) in the \apostle\ {suite}.  
For each simulated galaxy we identify the progenitor system at earlier times using the merger tree procedure described in \cite{2003MNRAS.338..903H}. We explicitly check that the progenitor ``tracks'' are smooth in position and mass (i.e., that the primary progenitor is accurately traced through time). For each system we define the ``accreted halo'' as the collection of star particles in the {\sc {subfind}} group (i.e., gravitationally bound to the host) whose FoF group at their time of formation is not the FoF group hosting the progenitor of the MW- or M~31-like galaxy at the same time. Typically, a satellite galaxy will become FoF-associated with its host before any substantial disturbances to the stellar and gas components of the satellite due to the massive host begin. Our sample therefore does not include stars formed in tidal tails after accretion, ``in-situ'' halo stars, or stars ejected from the central galaxy. We~note that our definition of ``accreted halo'' includes \emph{{all accreted stars}}.   
Many of these are located in the central regions of the object they were accreted by and might be observationally characterized as part of a bulge, rather than a halo, component.

%%%%%%%%%%%%%%%%%%%%%%%%%%%%%%%%%%%%%%%%%%
\section{Results}

\begin{figure}[H]
\centering
\includegraphics[width=.7\columnwidth]{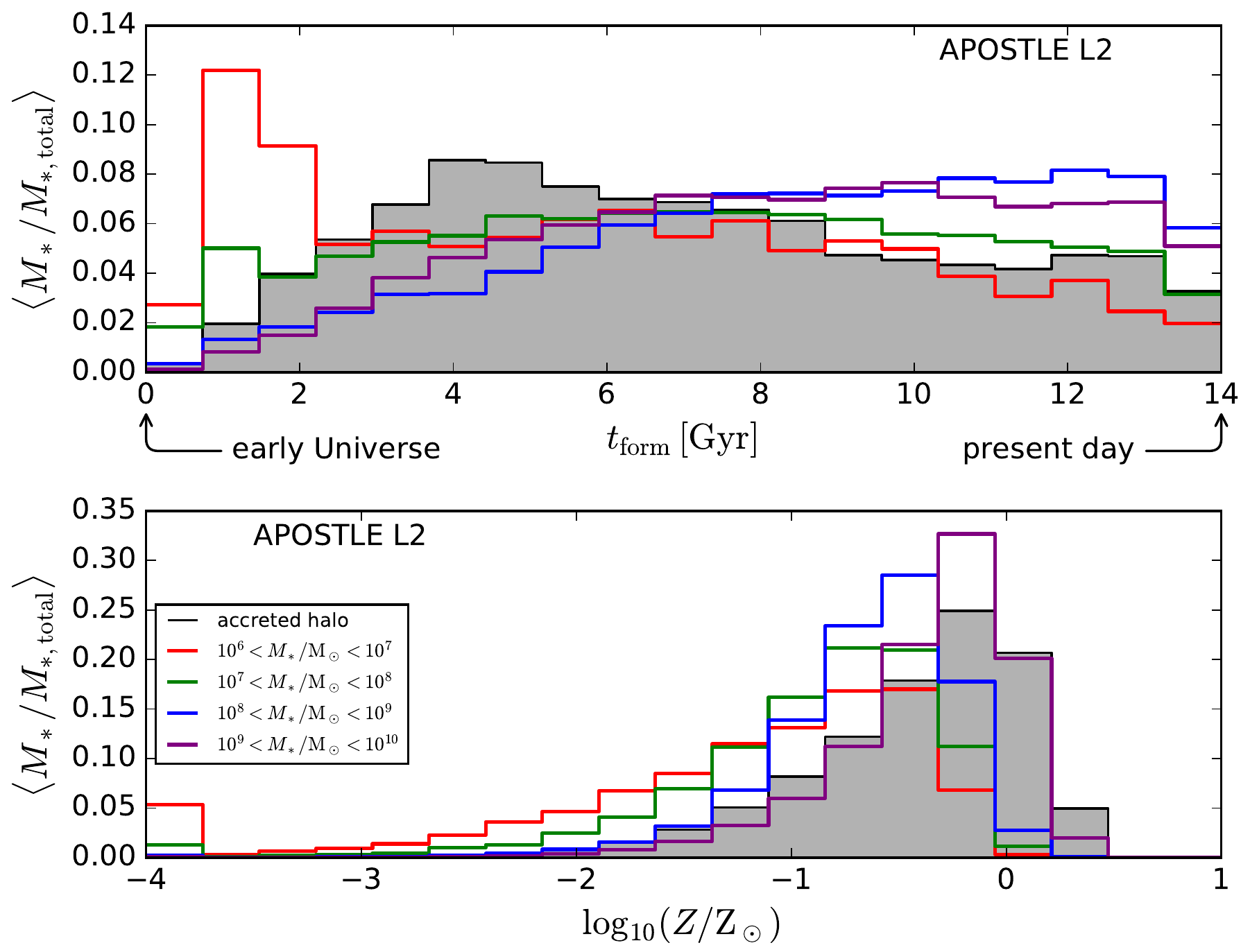}
\caption{The upper panel shows the formation time distribution for stars in the accreted halos of the $24$~Milky Way (MW)- and M~31-like galaxies from the \apostle\ simulation suite (filled histogram) at resolution level L2. The coloured lines show the same distribution for field (central) objects in $4$ consecutive 1-dex bins of stellar mass from $10^6$ to $10^{10}\,{\rm M}_\odot$ (red, green, blue, purple in order of increasing $M_\star$). The lower panel shows the metallicity distribution for the same classes of objects. The accreted halos have a metallicity distribution similar to that of $10^9$--$10^{10}\,{\rm M}_\odot$ field objects, but relatively older stellar populations. For all curves, star particles with metallicities $<$$-4$ contribute to the lowest metallicity bin shown. \label{fig1}}
\end{figure}

In Figure~\ref{fig1} we show the metallicity and formation time\footnote{the age of the Universe minus the age of the star} distributions of the accreted halo stars. The distributions are normalized to the total stellar mass of each system before they are combined, such that each system contributes equal weight to the distribution. The accreted halos as defined here are significantly more metal-rich (median $-0.31$) than recent estimates for the MW (median $-1.78$~\cite{2013ApJ...763...65A}), M~31 ($\lesssim$$-0.7$ \cite{2014ApJ...780..128I}), and other galaxies with similar stellar masses \cite{2017MNRAS.466.1491H}.~The chemical enrichment prescriptions adopted in the {\sc {eagle}}-Ref model---and therefore used in the \apostle\ simulations---result in a galaxy stellar mass--median metallicity relation offset to higher metallicities than observed \cite{2013ApJ...779..102K} for galaxies with $M_\star\lesssim10^8\,{\rm M}_\odot$. The disruption of these unusually metal-rich satellite galaxies unsurprisingly results in unusually metal-rich stellar haloes. This is fundamentally a shortcoming of the model. A direct comparison with the measurements cited above is further hindered by our selection of accreted particles, which inevitably includes many stars which would not usually be present in observed samples of ``halo stars'', especially toward the centre of each system. In the context of our analysis below, these differences are of limited concern since our argument concerns mainly relative---rather than absolute---metallicities.

With its tail of recently-formed stars, the formation time distribution may at first glance seem unusual: the MW stellar halo does not have such a population of young stars \cite{2016NatPh..12.1170C}. However, the M 31 stellar halo has a star formation history which, though it still peaks at ages of $5$--$13\,{\rm Gyr}$, has a clear tail to much younger ages \cite{2006ApJ...652..323B}. Furthermore, there is no attempt in the selection of systems for the \apostle\ simulations to choose halos with merger histories similar to the MW and M~31, or even the morphology of the galaxies: a few of the systems have had recent major mergers and still show clear morphological disturbances. These systems make the largest (but not the entire) contribution to the tail of young stars. How the \apostle\ sample of stellar haloes compares to the similarly-sized GHOSTS sample of stellar haloes  recently observed in detail \cite{2016MNRAS.457.1419M,2017MNRAS.466.1491H} (see also the compilation in \cite{2017ApJ...837L...8B}) is a topic we hope to pursue in a future contribution.

In Figure~\ref{fig1} we also show the formation time and metallicity distributions for other isolated (central) galaxies within the same \apostle\ simulation volumes, binned by stellar mass in 1~dex bins from $10^6$--$10^{10}\,{\rm M}_\odot$. The metallicity distribution of the accreted halos is roughly similar to that of the field objects in the highest mass bin ($M_\star\sim10^{9.5}$), but their formation time distribution is biased to earlier times (older ages). This is unsurprising: $10^{10}\,{\rm M}_\odot$ galaxies in the field are typically actively star forming up to the present day, whereas recently formed stars are excluded from the accreted halo sample which is dominated by disrupted ``quenched'' sattelites, as enforced by our selection process.

\begin{figure}[H]
\centering
\includegraphics[width=.7\columnwidth]{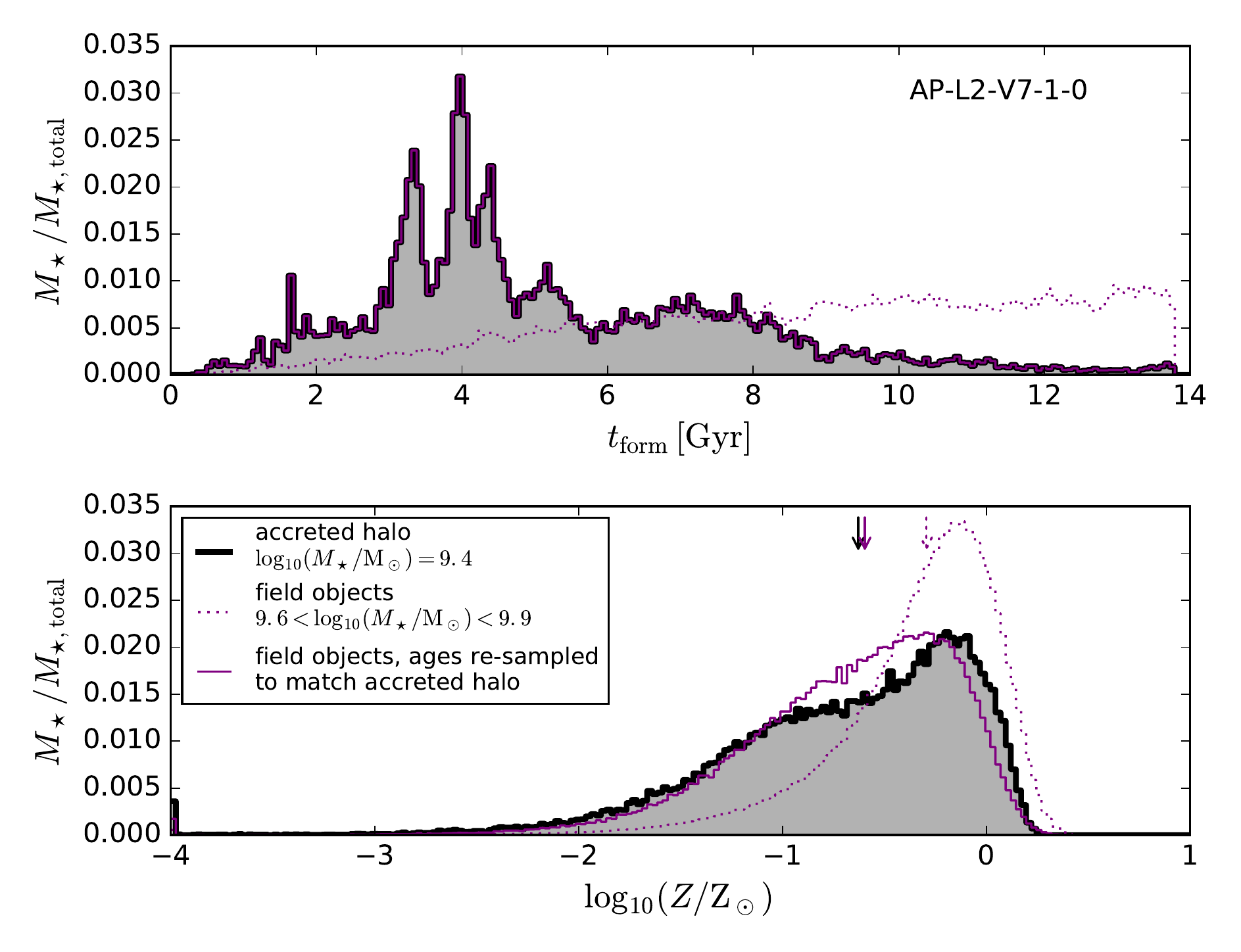}
\caption{The upper panel shows the formation time distribution for stars in the accreted halo of AP-L2-V7-1-0, one MW or M~31-like galaxy from the \apostle\ suite (filled histogram); the lower panel is similar but for the metallicity distribution. The dotted purple histogram shows the same distributions for field (central) objects which have stellar masses in the range $10^{9.6}$--$10^{9.9}\,{\rm M}_\odot$, i.e., $0.2$--$0.5$~dex more massive than the accreted halo of AP-L2-V7-1-0. The result of re-weighting the stellar populations of the field objects, weighted by the formation time distribution of the accreted halo, is shown with the solid purple histogram (by construction, the formation time distribution is then a perfect match). This enforced bias toward older stars in the field objects results in a re-sampled metallicity distribution which resembles much more closely that of the accreted halo. Arrows mark the median of the distribution with corresponding line style.\label{fig2}}
\end{figure}

The metallicity distributions in Figure~\ref{fig1} hint that the stellar populations in the accreted halos must be dominated by relatively massive accreted objects---the high median metallicity simply cannot be reached via the accretion of many low mass objects. We now explore this point further. For illustrative purposes, we use a single \apostle\ galaxy from the 7th volume, which we {label}\footnote{AP-[resolution level]-[volume number]-[FoF group number]-[subgroup number]} AP-L2-V7-1-0. This~galaxy was chosen ``by eye'' to be representative of the sample. The formation time and metallicity distributions of this galaxy are shown in Figure~\ref{fig2}. The total stellar mass of accreted halo stars in this system is $10^{9.4}\,{\rm M}_\odot$ (for comparison, the MW stellar halo mass is $\sim$$5.5\times10^8\,{\rm M}_\odot$ \cite{2016ARAnA..54..529B}, that of M~31 is $\sim$$1.5\times10^{10}\,{\rm M}_\odot$ \cite{2014ApJ...780..128I}; see also \cite{2017ApJ...837L...8B}). We also show the formation time and metallicity distributions for present-day dwarf galaxies in the field which have stellar masses slightly ($0.2$--$0.5$~dex; the choice of this particular interval is explained below) larger than this accreted halo. The stellar populations in these field galaxies are more metal-rich and younger than those in the accreted halo. However, if we re-weight the star particles in the same field objects such that the age distribution of the accreted halo is exactly matched, the resulting metallicity distribution is a close match to the metallicity distribution of the accreted halo, both in terms of the median (offset by less than $0.05$~dex, compared to $0.3$~dex before re-weighting) and the shape. 

\begin{figure}[H]
\centering
\includegraphics[width=.7\columnwidth]{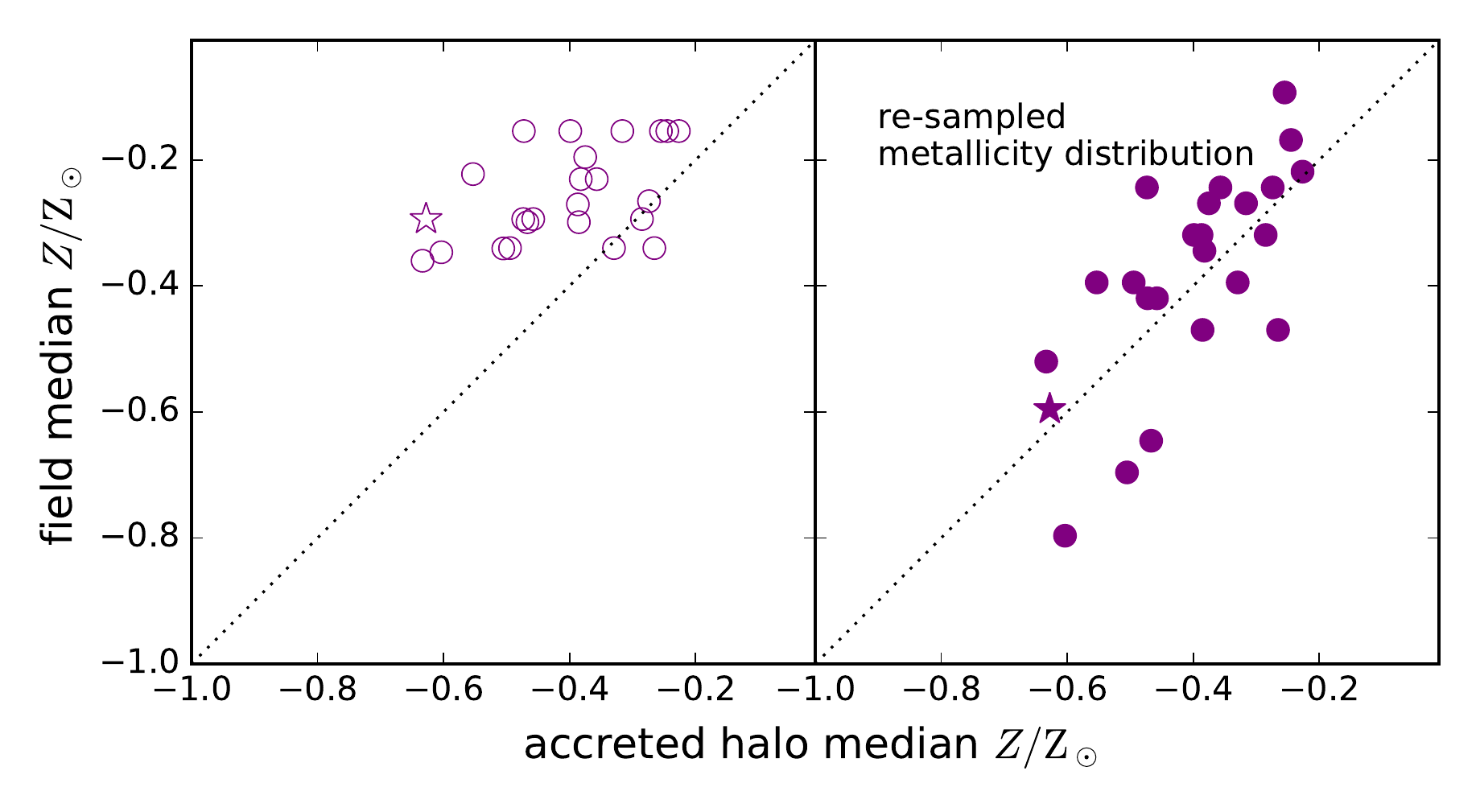}
\caption{Result of applying the stellar population re-weighting illustrated in Figure~\ref{fig2} to the $24$ MW- and M~31-like galaxies in the \apostle\ suite (AP-L2-V7-1-0, the example from Figure~\ref{fig2}, is marked with a star). In each case, the field objects in the stellar mass interval between $0.2$ and $0.5$~dex more massive than the stellar mass of the accreted halo are selected. Before the re-weighting (\textbf{left panel}), the field objects typically have a median metallicity greater than that of the corresponding accreted halo (the~dotted line indicates 1:1 agreement). After re-weighting (\textbf{right panel}), the medians of the metallicity distributions of the field objects and accreted halos agree, albeit with substantial scatter.\label{fig3}}
\end{figure}

In Figure~\ref{fig3} we show the result of applying the same process illustrated in Figure~\ref{fig2} to all $24$~accreted halos in our sample. We use the same offset of $0.2$--$0.5$~dex in stellar mass for all galaxies and show the median metallicity before and after re-weighting by the formation time distribution of the accreted halo. In most cases, the median of the re-weighted distribution approaches that of the accreted halo, though with significant scatter. The mass offset interval was chosen based on purely empirical considerations, by systematically exploring a range of possibilities covering the full mass range of field objects present in the simulations, and various widths for the interval. The $0.2$--$0.5$~dex window is the one which minimizes the scatter in the right panel of Figure~\ref{fig3}, without introducing a systematic offset from the line of 1:1 agreement.

%%%%%%%%%%%%%%%%%%%%%%%%%%%%%%%%%%%%%%%%%%
\section{Discussion}

The ``accreted halos'' from the \apostle\ simulation suite, as we have defined them here, should not be taken as direct detailed models of the MW, M~31, or other galactic stellar halos as defined observationally. They are, however, robust and internally self-consistent models of the assembly of such systems, and simultaneously of the nearby field objects which survive to the present day.

The above results suggest that the mass in the accreted halo of a galaxy is usually dominated by the disrupted content of one or a handful of relatively massive objects. If these had continued to grow and evolve in isolation instead of being accreted and destroyed, we would expect them to resemble present-day dwarf galaxies in the field with masses a factor of $\lesssim$$3$ greater than that of the accreted halo. Older stellar populations in relatively massive dwarf galaxies are roughly the surviving analogs of the most massive ``building blocks'' of stellar halos.

Though it seems that $1$--$2$ disrupted massive systems make up the bulk of most stellar halos, the remains of many lower-mass systems are also expected to be present. These are a nearly insignificant contribution (by mass) to the halo as a whole, but their signature may be detectable as a radial gradient---less massive systems are subject to weaker dynamical friction, and are destroyed at larger radii---and/or as overdense features such as shells or streams. The simulations and method used above offer a means of studying the link between the properties of such features and the types of objects which were destroyed to create them.

%%%%%%%%%%%%%%%%%%%%%%%%%%%%%%%%%%%%%%%%%%
%\section{Conclusions}

%This section is not mandatory, but can be added to the manuscript if the discussion is unusually long or complex.

%%%%%%%%%%%%%%%%%%%%%%%%%%%%%%%%%%%%%%%%%%
\vspace{6pt} 

%%%%%%%%%%%%%%%%%%%%%%%%%%%%%%%%%%%%%%%%%%
%% optional
%\supplementary{The following are available online at www.mdpi.com/link, Figure S1: title, Table S1: title, Video S1: title.}

%%%%%%%%%%%%%%%%%%%%%%%%%%%%%%%%%%%%%%%%%%
\acknowledgments{We thank the other members of the \apostle\ team, especially T. Sawala, A. Fattahi, M.~Schaller, C. Frenk, for their efforts in creating the simulations used in this work. We thank the {\sc {eagle}} simulation collaboration for providing the galaxy formation model, software and calibration. KAO thanks J. Helly for support in using the merger tree software. We thank the anonymous referees for their useful comments and suggestions. This work was supported by the Science and Technology Facilities Council (grant number ST/F001166/1). This~work used the DiRAC Data Centric system at Durham University, operated by the Institute for Computation Cosmology on behalf of the STFC DiRAC HPC Facility (www.dirac.ac.uk). This equipment was funded by BIS National E-infrastructure capital grant ST/K00042X/1, STFC capital grant ST/H008519/1, and STFC DiRAC Operations grant ST/K003267/1 and Durham University. DiRAC is part of the National E-Infrastructure. This~research has made use of NASA's Astrophysics Data System.}

%%%%%%%%%%%%%%%%%%%%%%%%%%%%%%%%%%%%%%%%%%
\authorcontributions{K.A.O., E.S. and J.F.N. conceived and designed the experiments; K.A.O. performed the experiments; K.A.O. and E.S. and J.F.N. analyzed the data; K.A.O. wrote the paper.}

%%%%%%%%%%%%%%%%%%%%%%%%%%%%%%%%%%%%%%%%%%
\conflictsofinterest{The authors declare no conflict of interest. The founding sponsors had no role in the design of the study; in the collection, analyses, or interpretation of data; in the writing of the manuscript, and in the decision to publish the results.} 

\end{document}